\documentclass[trackchanges]{aastex701}

\begin{document}

\title{The size of 3I/ATLAS from non-gravitational acceleration}

\author[orcid=0000-0000-0000-0001]{John C. Forbes}
\affiliation{School of Physical and Chemical Sciences--Te Kura Mat\=u, University of Canterbury, Christchurch 8140, New Zealand}
\email[show]{john.forbes@canterbury.ac.nz}  

\author[orcid=0009-0001-3617-8455]{Harvey Butler} 
\affiliation{School of Physical and Chemical Sciences--Te Kura Mat\=u, University of Canterbury, Christchurch 8140, New Zealand}
\email[show]{hbu41@uclive.ac.nz}


\begin{abstract}
The third macroscopic interstellar object detected in the solar system recently passed through perihelion, with the best-fitting models of its trajectory now featuring non-gravitational accelerations. We assess how much mass loss is required to produce plausible non-gravitational acceleration solutions and compare with estimates of the mass loss. We find that they are consistent when the nucleus of 3I/ATLAS is around 1 km in diameter. For a recent solution with a time lag in the acceleration from Eubanks et al, we find diameters between 820 meters and 1050 meters, assuming an outgassing asymmetry factor $\zeta=0.5$ and a density of the comet nucleus $\rho=0.5\ \mathrm{g}\ \mathrm{cm}^{-3}$. The limits on the diameter scale as $(\zeta/\rho)^{1/3}$. Substantial extrapolation is required in general to compare non-gravitational accelerations to mass loss rates, so reliable estimates of the mass loss rate at other stages of the comet's trajectory will substantially reduce the systematic uncertainty in this estimate. 
\end{abstract}


\section{Main}
As comets approach the Sun, higher temperatures cause sublimation of volatile ices, which depart the nucleus in the gas phase, carrying dust and ice grains along \citep{ahearn2012}. Asymmetries in the outgassing impart non‐gravitational accelerations (NGAs) on the nucleus via conservation of momentum \citep{whipple1950,Marsden1968}. Measurements of NGAs may therefore be combined with estimates of the mass loss rate to constrain the mass of the nucleus \citep[e.g.][]{sosa2009}. Importantly the acceleration is never measured at a single moment, but inferred from its cumulative effect on the astrometry, whereas mass loss rates are estimated at well-defined points in time. NGAs are parameterized as a smooth interpolation between two powerlaws in heliocentric distance, with parameters set according to the species under consideration \citep{Marsden1973, Krolikowska2023}, sometimes with the heliocentric distance evaluated at an offset time, $t+\Delta T$ \citep{Yeomans1989}

\begin{figure}[t!]
    \centering
    \includegraphics[width=0.9\linewidth]{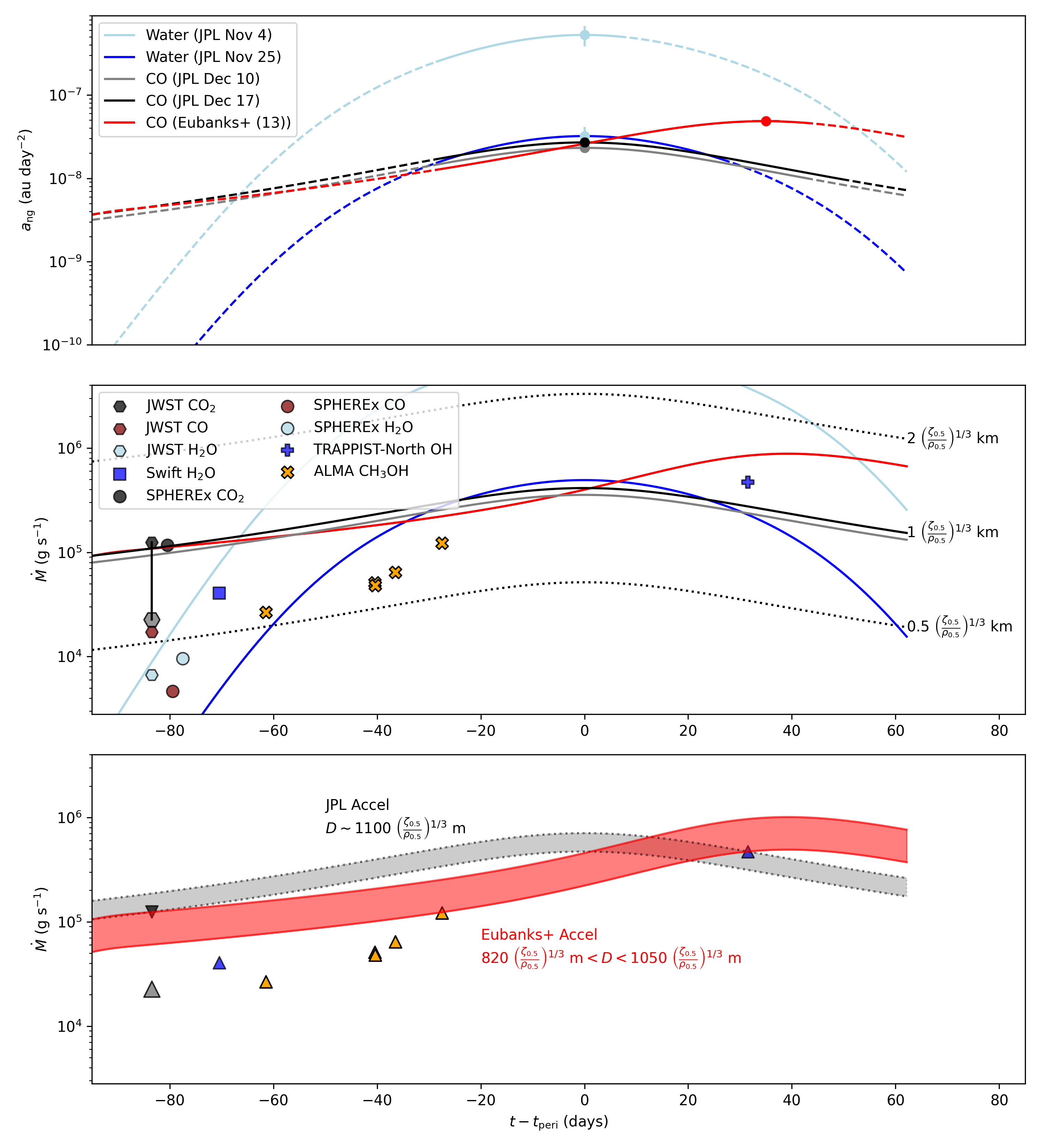}
    \caption{NGAs and their corresponding mass loss rates. {\bf Top} Non-gravitational acceleration from the JPL Small-Body Database or the final line of Table 1 in \citet{eubanks2025}. Reported errorbars are shown, and extrapolations into the future or prior to October are shown as dashed lines. {\bf Middle} NGAs from the top panel are translated into mass loss rates assuming 1 km diameters, with $\zeta=0.5$ and $\rho=0.5\ \mathrm{g}\ \mathrm{cm}^{-3}$. Additionally the JPL Dec 17 solution for other diameters are shown as dotted lines. Data points are from JWST \citep{Cordiner2025}, SPHEREx \citep{lisse2025b}, ALMA \citep{roth2025}, TRAPPIST-North \citep{jehin2025}, and Swift \citep{xing2025}. {\bf Bottom} CO$_2$ production rates measured by JWST are treated as upper and lower limits, and observations where CO$_2$ is inaccessible are shown as lower limits. Given these constraints, the range of $\dot{M}$ and hence diameters for two different reported NGA solutions is shown. The JPL solution is unable to simultaneously fit the upper limit from JWST and the lower limit from TRAPPIST-North.
    }
    \label{fig:thefigure}
\end{figure}


Over the past several months, the JPL Small-Body Database\footnote{\url{ssd.jpl.nasa.gov/tools/sbdb_lookup.html}} has reported significant NGAs for 3I/ATLAS. At first these were based on the \citet{Marsden1973} law for water sublimation, but have since been updated to use NGAs proportional to $(r/1\ \mathrm{au})^{-2}$ appropriate for CO$_2$ in the inner solar system. Several of these solutions are plotted in the top panel of Figure \ref{fig:thefigure}. We also show a solution from the table of \citet{eubanks2025} (its final row), which includes a time lag and data from spacecraft far from Earth. Since no NGAs were reported through October, we mark this region of the diagram with dashed lines to emphasize where the greatest constraints are likely arising. At the maximum value of $a_\mathrm{ng}$, we show a point with the reported errorbar, though it is much smaller than the systematic uncertainty.

To translate these NGAs into mass loss rates, $\dot{M}$, for comparison to such measurements, we use conservation of momentum,
\begin{equation}
    M |\vec{a}_\mathrm{ng}| = \left|\sum \dot{M}_i \vec{v}_i\right| \approx \zeta \dot{M} v_\mathrm{th}
\end{equation}
where the sum extends over all sources of outgassing on the comet's surface, each outgassing with velocity $\vec{v}_i$ and mass loss rate $\dot{M}_i$. We assume that these velocities are of order the thermal velocity, which we take to be $v_\mathrm{th} = 0.8\ \mathrm{km}\ \mathrm{s}^{-1} (r/1\ \mathrm{au})^{-0.5}$ following \citet{Cordiner2025}. The asymmetry is encapsulated in $\zeta
\lesssim1$, which is typically taken to be $\sim 0.5$ \citep{sosa2009}. Small values of $\zeta$ correspond to highly symmetric outgassing, while $\zeta \rightarrow 1$ means a single collimated jet. With these assumptions, we solve for $\dot{M}$, and plot a family of curves for each $a_\mathrm{ng}$ varying the mass. These curves are shown in the middle panel of Figure \ref{fig:thefigure}, where we retain the dependence on $\zeta$ and $\rho$ in our quoted diameters, normalizing each to a typical value, $\rho_{0.5} = \rho/(0.5\ \mathrm{g}\ \mathrm{cm}^{-3})$ and $\zeta_{0.5} = \zeta/0.5$. A value of $\zeta=0.5$ is in line with models presented in \citet{roth2025} and KCWI observations \citep{hoogendam2025}. Higher values of $\zeta$ are possible given the presence of prominent jets \citep{serraricart2025}, while lower values may be possible given the symmetry in the JWST observations \citep{Cordiner2025}.

The curves from each $a_\mathrm{ng}(r)$ solution are compared with observational estimates of the mass loss rate of various molecules in the middle panel of Figure \ref{fig:thefigure}. JWST's observation of strong CO$_2$ suggests that, at least at that epoch (6 August), the NGAs should be dominated by CO$_2$. However, only SPHEREx so far has reported additional CO$_2$-sensitive observations due to the inaccessibility of the line to ground-based observatories \citep{ootsubo2012,opitom2025}, and SPHEREx's observations were almost contemporaneous with JWST's \citep{lisse2025b}. Therefore in interpreting the diagram, one should think of the observations that cannot detect CO$_2$ as conservative lower limits of $\dot{M}$ at that time. Nonetheless, these lower limits may be constraining. We do not include dust mass loss rates, as reported for instance by \citet{jewitt2025b}, since the dust may be ejected at far slower speeds than the gas \citep[e.g.][]{thomas2015}. Note also that for the JWST CO$_2$ observation, we include both the asymptotic value of the production rate far from the nucleus (black) and the innermost value (gray point) since some CO$_2$ gas may come from sublimation of grains in the coma \citep{lisse2025b}. However, \citet{Cordiner2025} argue that the difference between these two points comes from optical depth effects.

In the lower panel, we show the effect of enforcing these upper and lower limits. With the \citet{eubanks2025} acceleration model, they result in diameters between $820\ (\zeta_{0.5}/\rho_{0.5})^{1/3}$ meters and $1050\ (\zeta_{0.5}/\rho_{0.5})^{1/3}$ meters, with both the TRAPPIST-North and ALMA datapoints constraining the lower end, and JWST constraining the upper end. Meanwhile, the JPL NGA law is unable to simultaneously match these same limits. It is possible that certain simplifying assumptions like $\zeta=$const., $M=$const., or the NGA being dominated by a single molecular species, could be relaxed to relieve this tension. Another possibility apparently supported by the astrometry is the time-delayed acceleration suggested by \citet{eubanks2025}. Our reported diameters are slightly larger than their estimate, likely due to differences in assumed ejection velocities and choice of the particular NGA solution. Regardless, our diameter estimates are comfortably within the upper limits from {\em HST} \citep{jewitt2025}.

\begin{acknowledgments}
We thank Michele Bannister and Chris Lintott for helpful comments.
\end{acknowledgments}

\begin{contribution}

JCF produced the figure and edited the manuscript; HB wrote early versions of the code and text.


\end{contribution}

\bibliography{refs}
\bibliographystyle{aasjournalv7}



\end{document}